\documentclass{aastex631}

\usepackage{subfigure}
\usepackage{amsmath}
\usepackage {booktabs}

\begin{document}

\title{PNet -- A Deep Learning Based Photometry and Astrometry Bayesian Framework}

\correspondingauthor{Peng Jia}
\email{robinmartin20@gmail.com}

\author{Rui Sun}
\affiliation{College of Electronic Information and Optical Engineering, Taiyuan University of Technology, Taiyuan, 030024, China}

\author[0000-0001-6623-0931]{Peng Jia}
\affiliation{College of Electronic Information and Optical Engineering, Taiyuan University of Technology, Taiyuan, 030024, China}
\affiliation{Peng Cheng Lab, Shenzhen, 518066, China}
\affiliation{Department of Physics, Durham University, DH1 3LE, UK}

\author{Yongyang Sun}
\affiliation{College of Electronic Information and Optical Engineering, Taiyuan University of Technology, Taiyuan, 030024, China}

\author{Zhimin Yang}
\affiliation{College of Electronic Information and Optical Engineering, Taiyuan University of Technology, Taiyuan, 030024, China}

\author{Qiang Liu}
\affiliation{College of Electronic Information and Optical Engineering, Taiyuan University of Technology, Taiyuan, 030024, China}

\author{Hongyan Wei}
\affiliation{College of Electronic Information and Optical Engineering, Taiyuan University of Technology, Taiyuan, 030024, China}


\begin{abstract}
Time domain astronomy has emerged as a vibrant research field in recent years, focusing on celestial objects that exhibit variable magnitudes or positions. Given the urgency of conducting follow-up observations for such objects, the development of an algorithm capable of detecting them and determining their magnitudes and positions has become imperative. Leveraging the advancements in deep neural networks, we present the PNet, an end-to-end framework designed not only to detect celestial objects and extract their magnitudes and positions but also to estimate photometry uncertainty. The PNet comprises two essential steps. Firstly, it detects stars and retrieves their positions, magnitudes, and calibrated magnitudes. Subsequently, in the second phase, the PNet estimates the uncertainty associated with the photometry results, serving as a valuable reference for the light curve classification algorithm. Our algorithm has been tested using both simulated and real observation data, demonstrating the PNet's ability to deliver consistent and reliable outcomes. Integration of the PNet into data processing pipelines for time-domain astronomy holds significant potential for enhancing response speed and improving the detection capabilities for celestial objects with variable positions and magnitudes.
\end{abstract}

% \keywords{Classical Novae (251) --- Ultraviolet astronomy(1736) --- History of astronomy(1868) --- Interdisciplinary astronomy(804)}


\section{Introduction} \label{sec:intro}
In recent years, time-domain astronomy has emerged as an active research field. With the availability of telescopes possessing a wide field of view and high image quality, it has become feasible to capture images of celestial objects at regular intervals, yielding a substantial amount of observational data on a daily basis. Among this vast dataset, there are numerous celestial objects that necessitate frequent or immediate follow-up observations, such as tidal disruption events, near-earth objects, super flares, and microlensing events. Consequently, there is a pressing need to develop an algorithm capable of swiftly detecting these events. Since these events primarily involve changes in the positions and magnitudes of celestial objects, the algorithm must possess the capability to detect celestial objects and conduct precise photometry and astrometry measurements. Furthermore, given that images of celestial objects are susceptible to various sources of noise, the algorithm should also be able to estimate the uncertainties associated with the photometry results, enabling further in-depth analysis.\\

Numerous pipelines have been proposed to meet these requirements, typically comprising the following key steps:
\begin{itemize}
    \item Target detection: The positions of potential celestial object candidates are determined.
    \item Target classification: True celestial objects are identified from the pool of candidates, and further categorized into different types.
    \item Target information extraction: Magnitudes, positions, and distributions of the celestial objects are obtained.
\end{itemize}

Previous studies have introduced a variety of algorithms to establish the conventional data processing pipeline. Typically, target detection algorithms such as SExtractor or simplexy have been utilized to identify potential celestial objects from the original observational images \citep{lang2010astrometry, bertin1996sextractor}. Subsequently, these identified targets undergo classification algorithms that aim to distinguish true celestial objects from the candidate pool \citep{cabrera2017deep, duev2019real, jia2019optical, turpin2020vetting, agarwal2020fetch}. The resulting information regarding these celestial objects is then processed through photometry, astrometry, morphology classification, or segmentation algorithms \citep{khramtsov2019deep, boucaud2020photometry, hausen2020morpheus, dominguez2022sdss, casetti2023star}. However, the classical data processing pipeline follows a sequential structure, wherein all the processes are executed in sequence. Consequently, the overall performance of the data processing pipeline is limited by the performance of each individual algorithm used. For example, if celestial objects are not detected by the source detection algorithm, it becomes impossible to extract information related to those targets. It should be noted that contemporary source detection algorithms possess numerous adjustable parameters, requiring the expertise of experienced scientists to properly set them. In addition, since the detection results are sensitive to environmental conditions, frequent human intervention is required to obtain effective results. Therefore, it is necessary to develop an end-to-end framework that could not only detect celestial objects, but also extract their information automatically and robustly.\\

Deep neural network-based algorithms for celestial object detection have attracted considerable attention in recent years. One key advantage is that these algorithms enable end-to-end learning, allowing the neural network to directly acquire the ability to detect celestial objects. Different tasks can be addressed by designing and deploying deep neural networks with specific architectures \citep{ren2015faster, ge2021yolox, liu2021swin}. In this study, our focus is on detecting point-like celestial objects in sparse star fields and extracting their positions and magnitudes, as this is a crucial prerequisite for studying such objects in time-domain astronomy. It is worth noting that extended targets and dense star fields (the distance between stars are less than 2 times of the full width half magnitude of the point spread function) may benefit from multicolour images and other relevant neural networks for better detection and classification \citep{gonzalez2018galaxy, farias2020mask, cheng2021galaxy, jia2022detection, jia2023deep, yu2022investigations,andrew2023galaxy} , or from methods specifically designed for dense star fields \citep{liu2021variational,Hansen2022}. Moreover, transients, such as supernovae in galaxies, can be further processed using image difference-based methods or techniques developed based on temporal sequences of images \citep{wright2015machine, kessler2015difference, zackay2016proper, sanchez2019machine, mong2020machine, gomez2020classifying, hu2022image, makhlouf2022train}. A previous study by \citet{jia2020detection} has introduced a Faster-RCNN based framework for detecting point-like celestial objects, which was successfully applied to images captured by wide field optical telescopes and the Lobster-Eye telescope \citep{jia2023target}. However, in real applications, there are three key challenges that need to be addressed:\\
1. The current framework does not provide apparent magnitudes as part of its output. Apparent magnitudes play a crucial role in various tasks, such as exoplanet observations or studying super-flares from stars. Therefore, it is essential to integrate a photometry algorithm into the detection framework to accurately estimate the apparent magnitudes of the celestial targets.\\
2. Contemporary deep neural network (DNN) based target detection algorithms define the positions of celestial objects using bounding boxes, which are rectangular boxes that approximate the shape of the objects. However, to match these celestial objects with catalogs and perform further analysis, we need to determine the precise centers of the celestial object images. Hence, it is necessary to incorporate an astrometry method into the detection framework to accurately estimate the centers of different celestial objects.\\
3. The current framework lacks the ability to estimate uncertainties associated with magnitude estimation. Since most neural networks provide point estimates for a given input, they directly output regression values without accounting for uncertainties. However, uncertainties are vital for subsequent tasks, such as light curve classification. Therefore, it is crucial to develop a method that can estimate the uncertainties introduced by magnitude estimations.\\

To enhance the suitability of our Faster-RCNN based astronomical detection algorithm for integration into time domain astronomy data processing pipelines, further improvements are necessary. In this study, we present our endeavour to develop a novel framework called the deep learning based photometry and astrometry Bayesian Neural Network (PNet). The PNet includes an advanced architecture that excels in detecting, performing photometry, and carrying out astrometry for point-like celestial objects. Notably, the PNet leverages the Bayesian Neural Network (BNN) for estimating both the photometry results and their associated uncertainties. The subsequent sections of this paper will delve into various aspects of our work. Section 2 will discuss the properties of the data and the methods employed for data reprocessing. In Section 3, we will present the structure of the PNet. In Section 4, we will assess the PNet's performance using both simulated data and real observation data. And these results will be compared to those obtained using the SExtractor \citep{bertin1996sextractor} to show the advantageous of the PNet. Finally, in Section 5, we will conclude our findings and outline our future research directions.\\

\section{The Data} \label{sec:data}
In this paper, we train and evaluate the performance of our framework using simulated and real observation data. Simulated data allows us to have control over the observation conditions, enabling a more precise assessment of our framework's performance. On the other hand, real observation data encompass various unknown factors that better reflect the actual performance of our framework. To generate the simulated data, we use Skymaker, a widely used tool for generating synthetic images based on specified observation conditions \citep{bertin2009skymaker}. Skymaker generates simulated images by considering parameters such as point spread functions, noise levels, and input star catalogs, which are obtained by the Stuff for galaxies and manually generated catalogs for stars. The distribution of photons emitted by celestial objects follows a Poisson distribution, with the point spread function serving as the prior distribution function. Additionally, the simulation includes the generation of Poisson-distributed sky background photons. To account for readout noise, Gaussian noise is simulated, and effects like blooming/bleeding are also considered in Skymaker. By carefully controlling parameters in the Skymaker, we ensure that the PSF size and noise level closely resemble those of real observation data, thus facilitating a thorough investigation of our framework's performance.\\

The real observation data employed in this paper is derived from the Sloan Digital Sky Survey (SDSS) DR17 \citep{Abdurro’uf_2022}. The SDSS data is collected using a wide-field 2.5m telescope \citep{Gunn_2006} located at the Apache Point Observatory in New Mexico \citep{York_2000}. This data undergoes meticulous processing through a specialized data processing pipeline \citep{lupton2005sdss}, which includes precise astrometric calibration \citep{Pier_2003} using the USNO CCD Astrograph Catalog (UCAC) \citep{Zacharias_2000}, as well as photometric calibration \citep{Padmanabhan_2008} with the aid of numerous standard stars \citep{Smith_2002}. Consequently, the data and catalogue obtained from the SDSS serve as reliable references for both training and testing purposes. The SDSS employs five filters, namely u, g, r, i, and z filters, which correspond to central wavelengths of 3543, 4770, 6231, 7625, and 9134 \AA, respectively \footnote{\url{https://skyserver.sdss.org/dr1/en/proj/advanced/color/sdssfilters.asp}}. In this study, we focus on images obtained using the g-band filter due to their relatively higher signal-to-noise ratio. Additionally, the parameters $run$, $camcol$, $field$, and $rerun$ define the observation period of the telescope, the camera column number, the area number on the camera, and the version number during data processing, respectively. To ensure the generalization of our algorithm across different sky zones and time periods, we avoid specifying specific values for the first three parameters. However, we do specify the use of data processing process with version number 301 for the $rerun$ parameter.\\

\begin{figure}
    \plotone{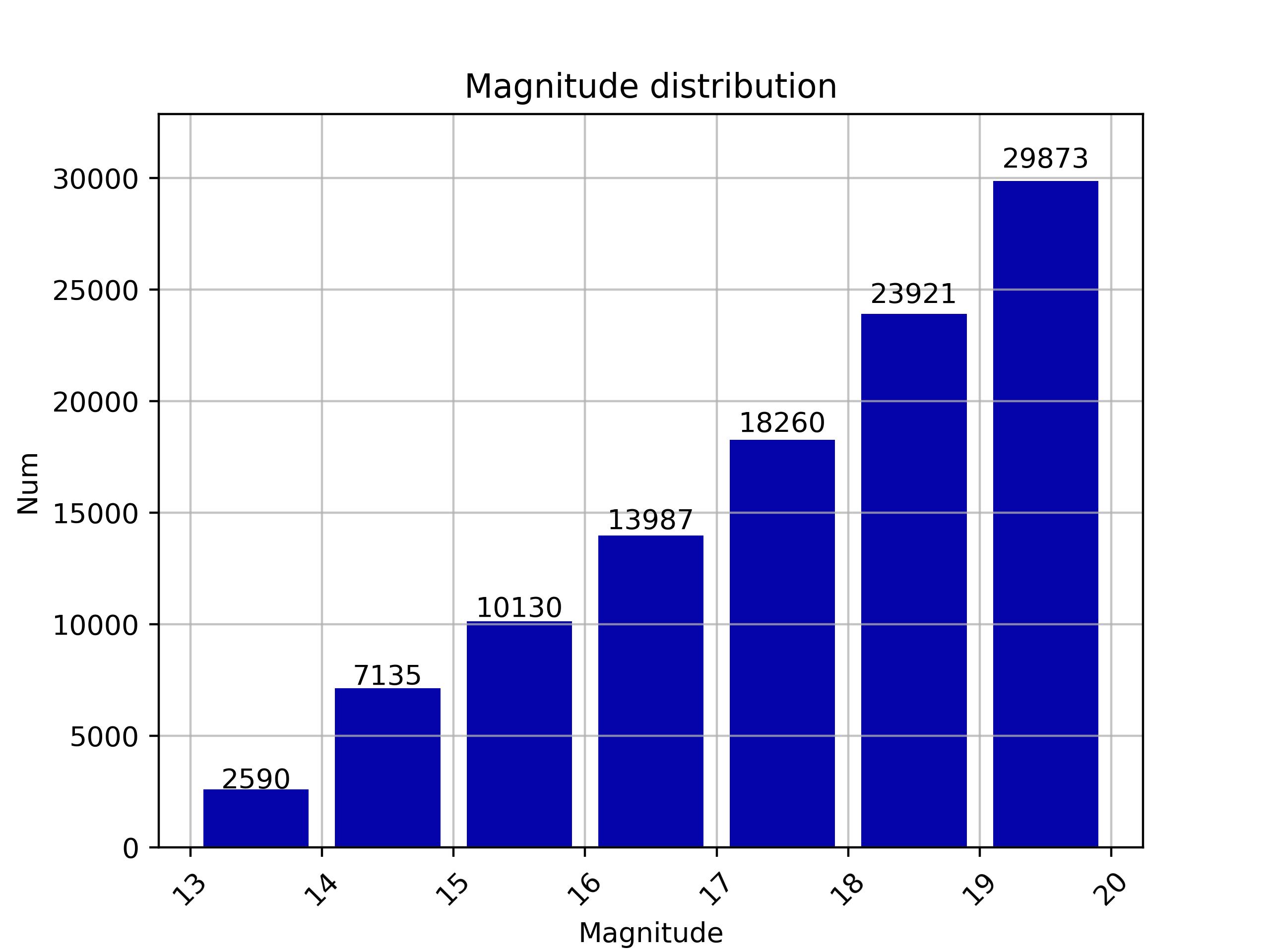}
    \caption{The histogram of stars with different apparent magnitude in the real data set.}
    \label{fig:sdssmagstats}
\end{figure}

Given that our framework is designed to specifically detect point-like celestial objects, the dataset exclusively consists of such objects. The magnitudes and positions of these point-like celestial objects are determined as regression values within the framework. The resulting positions are provided in camera coordinates, using pixels as a unit of measurement, while the magnitudes are represented as flux $f$. To calculate the apparent magnitude, we utilize the equation defined in \citep{Stoughton_2002} as follows:
\begin{equation}
    mag = 22.5-2.5\log_{10} f,
    \label{equ:mag}
\end{equation}
The apparent magnitude ($mag$) is related to the flux ($f$) according to the equation provided. In this study, we adopt a magnitude zero point of 22.5 and estimate the magnitudes of stars within the range of 13 to 20. Figure~\ref{fig:sdssmagstats} illustrates the histogram depicting the distribution of stars by different magnitudes in the real data set. The observed distribution aligns with our experience.\\

Taking into account the impact of input image size on GPU memory requirements, we divide the original SDSS images into patches with size of $512\times512$ pixels. This approach helps reduce the hardware demands. Additionally, we apply certain criteria to remove specific stars that would otherwise necessitate additional processing steps. These criteria include:
\begin{itemize}
    \item Stars located at the image's edge within a 10-pixel distance.
    \item Stars situated near galaxies or other objects.
    \item Stars positioned in close proximity to one another within a 10-pixel range.
\end{itemize}

\section{The Method} \label{sec:meth}
\begin{figure}[ht!]
    \plotone{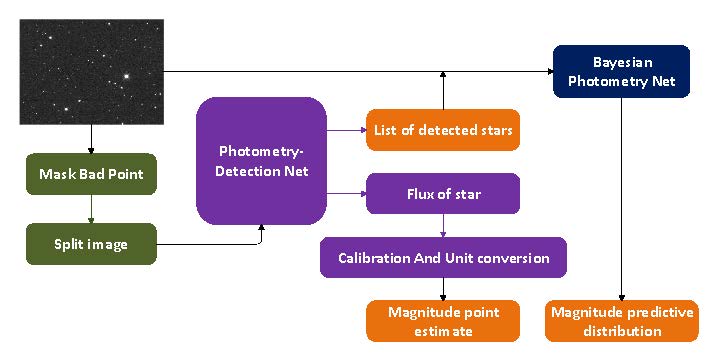}
    \caption{The flowchart of the PNet proposed in this paper. The green block indicates the data preprocessing part, the purple block indicates the Photometry-Detection Net part, the blue block indicates the Bayesian Photometry Neural Network part, and the orange block indicates the output of the whole algorithm.}
    \label{fig:flowchart}
\end{figure}

The flowchart in Figure~\ref{fig:flowchart} illustrates the structure of the proposed framework. Initially, we identify and mask out any defective pixels present in the original image. Following that, we divide the image into patches with size of $512\times512$ pixels. Subsequently, we employ the Photometry Detection Net to detect point-like targets within these image patches and conduct photometry to determine the flux of these identified targets. Using the obtained flux values, we perform photometry calibration to derive the magnitudes of the stars. Lastly, we utilize the Bayesian Photometry Neural Network (BPNN) to re-evaluate the magnitudes of the stars and estimate the associated uncertainty in the magnitude measurements. The decision to separate the star detection and the BNN for magnitude measurement is based on the DNN's extensive parameterization, which necessitates multiple sampling during the prediction phase, thereby requiring significant computational resources. In this section, we will first introduce several performance evaluation metrics. We will then provide a concise overview of the Bayesian Neural Network (BNN) and subsequently discuss the implementation details and training procedures of the framework. All the neural networks described in this paper are constructed using PyTorch and executed on a computer equipped with a GTX 3090Ti GPU.\\

\subsection{The Performance Evaluation Criterion} \label{subsec:criterion}
Choosing appropriate performance evaluation criteria is crucial for properly assessing the performance of a framework. It is essential to select evaluation criteria that align with the objectives of developing the framework. Our algorithm focuses on four key aspects: detection of point-like targets, regression of their positions and magnitudes, and estimation of photometric uncertainties. Hence, the following performance evaluation criteria have been selected for our algorithm:
\begin{itemize}
    \item The recall rate and the precision rate, and mean Average Precision (mAP) are chosen to evaluate the performance of target detection results.
    \item The astrometry accuracy in pixels is used to assess the accuracy of the astrometry results.
    \item The photometry accuracy in magnitudes is used to evaluate the accuracy of the photometry results.
    \item The outlier fraction ($\eta$), Normalized Median Absolute Deviation (NMAD), Median Absolute Deviation (MAD), and the mean value of the photometry results $1 \sigma$ ($\bar{E}$) are utilized to evaluate the uncertainty of the photometry results.
\end{itemize}

In the following, we will provide a detailed description of the aforementioned performance evaluation criteria. When evaluating the detection results, we consider four possible scenarios:
\begin{itemize}
    \item True Positive (TP): This occurs when a point-like celestial object is correctly identified as a point-like celestial object.
    \item True Negative (TN): This occurs when targets other than point-like celestial objects are correctly identified as non-point-like celestial objects.
    \item False Positive (FP): This occurs when targets other than point-like celestial objects are wrongly identified as point-like celestial objects.
    \item False Negative (FN): This occurs when point-like celestial objects are wrongly identified as non-point-like celestial objects.
\end{itemize}

Given that our framework is capable of directly outputting the center coordinates of the point-like celestial objects, we assess the detection results by calculating the Euclidean distance between the predicted position and the corresponding position in the label. The Euclidean distance can be defined using the following equation~\ref{equ:euc}:
\raggedbottom
\begin{equation}
    EuclideanDistance = \sqrt{\sum_{i=1}^{n}{(x_i - y_i)^2}},
    \label{equ:euc}
\end{equation}
the variables $x_i$ represent the predicted positions, $y_i$ represent the label positions, and $n$ represents the number of coordinates used to describe the positions (which is 2 in this paper). If the Euclidean distance between the predicted positions and the label positions is below a certain threshold and the classification result is in accordance with the label, it will be considered true positive (TP) or true negative (TN). Otherwise, it will be classified as a false positive (FP) or false negative (FN).\\

Based on the aforementioned definition, we can assess the performance of our framework in target detection by utilizing the Precision and Recall metrics, as defined in the following Equation~\ref{equ:PR}:
\begin{equation}
    \begin{aligned}
        Precision &= \frac{TP}{TP + FP},\\
        Recall &= \frac{TP}{TP + TN}.
    \end{aligned}
    \label{equ:PR}
\end{equation}
Precision and Recall are widely used metrics for evaluating target detection results. Precision represents the percentage of true positives among all positive detection results, indicating the performance of the detection algorithm in minimizing false alarms. Recall, on the other hand, represents the percentage of true positives among all actual targets, describing the ability of the detection algorithm to identify all positive instances. Precision and Recall are both influenced by the chosen detection threshold. A higher threshold leads to higher precision, but lower recall, and vice versa. To comprehensively evaluate the performance of a detection algorithm, we can vary the detection threshold and generate a precision-recall curve (P-R curve). The area under the P-R curve is known as the average precision (AP). The mean average precision (mAP) is calculated by averaging the AP values across all categories, providing an overall assessment of the detection algorithm's performance. In this paper, the mAP is obtained using Equation~\ref{equ:map}, where $n$ represents the number of categories (which is 1 in this case). Furthermore, it is worth noting that the astrometry accuracy can be measured by calculating the distance between the predicted position and the corresponding position in the label for all true positive (TP) detection results.
\begin{equation}
    mAP = \frac{\sum{AP}}{n}
    \label{equ:map}
\end{equation}

The Bayesian neural network is employed to estimate the uncertainty inherent in the data. When a star image is fed into the Bayesian neural network, it generates a predictive posterior distribution for the star's magnitude, along with the mean value of that distribution. The mean and standard deviation of this distribution, which cannot be directly obtained, are typically estimated through multiple Monte Carlo sampling iterations. The standard deviation $\sigma_{mag}$ of this distribution can be interpreted as a quantitative measure of uncertainty. Consequently, we consider two aspects of the output: the deviation of the true value from the mean of the predicted distribution and the level of uncertainty in the prediction results. To evaluate the deviation of the true value from the mean, we employ the relative error $RelativeError$ and the absolute error $\delta_{mag}$, defined in Equation~\ref{equ:nmad1}:
\begin{equation}
    \begin{aligned}
        \delta_{mag} &= mag_{true} - mag_{pred},\\
        RelativeError &= \frac{\delta_{mag}}{1 + mag_{true}}.
    \end{aligned}
    \label{equ:nmad1}
\end{equation}
By considering the relative error for all targets, we can determine the fraction of outliers ($\eta$) by establishing a threshold value for the relative error and calculating the proportion of prediction results exceeding this threshold ($\sigma_{mag}$ in this study). Additionally, we employ the normalized median absolute deviation (NMAD) to evaluate the relative errors obtained. The median absolute deviation (MAD) is a statistical measure that characterizes the sample bias of one-dimensional numerical data. To obtain the NMAD, we normalize the MAD by multiplying it by a factor of 1.4826 \citep{rousseeuw1993alternatives}, as demonstrated in Equation~\ref{equ:nmad2},
\begin{equation}
    \begin{aligned}
        \delta_{NMAD} &= 1.48 \times median(\frac{abs(\delta_{mag} – median(\delta_{mag}))}{1 + mag_{true}}).
    \end{aligned}
    \label{equ:nmad2}
\end{equation}
The absolute error is assessed through the Mean Absolute Error (MAE), which represents the average of the absolute differences between the mean of the predictive posterior distribution and the corresponding true value. This metric provides a visual indication of the error level. Regarding uncertainty, we initially estimate the magnitude distribution for each star using the Bayesian neural network. We then compute the mean value of these distributions, denoted as $\bar{E}$, which serves as a reference indicator for the uncertainty distribution.\\

\subsection{The Principles of the Bayesian Photometry Neural Network} \label{subsec:prinbnn}
The Bayesian photometry neural network is utilized to estimate both the magnitude of stars and the associated photometry uncertainties. In traditional neural networks, the weights are fixed, leading to a lack of uncertainty estimation and excessive confidence in the predicted results. To address this issue, Bayesian neural networks employ the Variational Bayes (VB) method \citep{pmlr-v37-blundell15} to introduce uncertainty into the network weights. However, before delving into the methods for capturing uncertainty, it is crucial to comprehend the origins of uncertainty.\\

In the field of machine learning, two main types of uncertainty are commonly recognized: aleatoric uncertainty (also known as data uncertainty) and epistemic uncertainty (also known as model uncertainty) \citep{KIUREGHIAN2009105}. Aleatoric uncertainty (AU) stems from the inherent noise present in the dataset itself \citep{gal2016uncertainty}. Since this noise is natural and unpredictable, aleatoric uncertainty cannot be eliminated. On the other hand, epistemic uncertainty (EU) arises from insufficient training of the network, resulting in a lack of knowledge about the system's behavior. In principle, this uncertainty can be reduced as the training data approaches infinity \citep{HORA1996217}. As mentioned earlier, the predictive uncertainty (PU) we aim to capture can be expressed as the combination of AU and EU \citep{ABDAR2021243}, as illustrated in Equation~\ref{equ:pu}:
\begin{equation}
    PU = AU + EU
    \label{equ:pu}
\end{equation}

By defining the $PU$, we can establish the underlying principle of the Bayesian Photometry Neural Network (BPNN). Initially, we define the weights of the BPNN as $\omega \in \Omega$, where $\Omega$ represents the parameter space of the BPNN. The training dataset is denoted as D, and within this dataset, we have data pairs X and Y. Similarly, in the test dataset, we have data pairs x and y. The distribution of weights $\omega$ learned by the network from the dataset D is represented as $p(\omega \vert D)$. Additionally, $p(y \vert x, \omega)$ signifies the probability that the neural network yields output y when given input x and weight $\omega$. Lastly, $p(\omega)$ represents the prior weight distribution of the network. With these definitions, we can derive the probability distribution of the output y given the input x when the neural network is trained using dataset D, as illustrated in Equation~\ref{equ:pxd}:
\begin{equation}
    p(y \vert x,D) = \int p(y \vert x,\omega)p(\omega \vert D) d\omega.
    \label{equ:pxd}
\end{equation} 

The prediction uncertainty captured by the BPNN is represented by $p(y|x,D)$. However, obtaining $p(\omega \vert D)$ through analytical calculations is challenging in real applications, often requiring approximation methods to perform the inference task \citep{PhysRevD.102.103509}. In this study, we employ variational inference to approximate the solution for $p(\omega \vert D)$. Initially, we assume a variational distribution $q(\omega \vert \theta)$, where $\theta$ denotes a set of variational parameters. Subsequently, we calculate the Kullback-Leibler (KL) divergence between the variational distributions $q(\omega \vert \theta)$ and $p(\omega \vert D)$. Finally, we determine a set of variational parameters $\theta^\ast$ that minimizes the KL divergence, as shown in Equation~\ref{equ:thetamin}:
\begin{equation}
    \begin{aligned}
        \theta^\ast &= \mathop{\arg\min}\limits_{\theta} KL\left[q(\omega \vert \theta) \Vert p(\omega \vert D)\right] \\
        &= \mathop{\arg\min}\limits_{\theta} \int q(\omega \vert \theta) \ln{\frac{q(\omega \vert \theta)}{p(\omega \vert D)}} d\omega
    \end{aligned}
    \label{equ:thetamin}
\end{equation}
Since $p(\omega \vert D)$ cannot be obtained analytically, we further introduce the Bayesian formula below:
\begin{equation}
    p(\omega \vert D) = \frac{p(D \vert \omega)p(\omega)}{p(D)}.
    \label{equ:bayes}
\end{equation}
We could obtain the KL divergence according to Equation~\ref{equ:thetamin} and Equation~\ref{equ:bayes}, as shown in Equation~\ref{equ:kl}:
\begin{equation}
    KL\left[q(\omega \vert \theta) \Vert p(\omega \vert D)\right] = \ln{p(D)} + KL\left[q(\omega \vert \theta) \Vert p(\omega)\right] - \int q(\omega \vert \theta)\ln{p(D \vert \omega)} d\omega.
    \label{equ:kl}
\end{equation}
Since $\ln{p(D)}$ is only related to properties of data, we could minimize the KL divergence through minimizing the following Equation~\ref{equ:FDtheta}:
\begin{equation}
    \begin{aligned}
        F(D, \theta) &= KL\left[q(\omega \vert \theta) \Vert p(\omega)\right] - \int q(\omega \vert \theta)\ln{p(D \vert \omega)} d\omega \\
        &= E_{q(\omega \vert \theta)} \left[\ln{q(\omega \vert \theta) - \ln{p(D, \omega)}} \right],
    \end{aligned}
    \label{equ:FDtheta}
\end{equation}
It is important to note that the term $F(D,\theta)$ in Equation~\ref{equ:FDtheta} corresponds to the negative value of the Evidence Lower Bound (ELBO), as discussed in \citet{doi:10.1080/01621459.2017.1285773}. By employing Equation~\ref{equ:FDtheta}, we can transform the analytically challenging calculation into a practical optimization problem for the variational parameter $\theta$. For a comprehensive understanding of the approximation methodology, we refer readers to the work of \citet{PhysRevD.102.103509}, while providing a concise overview here.\\

To begin, let us assume that we can obtain a set of variational parameters $\hat{\theta}$ that minimizes $F(D, \theta)$ through an optimization process. In this study, we perform such optimization using the backpropagation algorithm \citep{Rumelhart1986} and employ the Adam gradient descent algorithm \citep{Kingma2014AdamAM}. With the obtained $\hat{\theta}$, we can derive the predictive distribution $q_{\hat{\theta}}$. By combining this with Equation~\ref{equ:pxd}, we can determine the predictive distribution of the output variable y given the input x:
\begin{equation}
    q_{\hat{\theta}}(y \vert x) = \int p(y \vert x, \omega)q(\omega \vert \hat{\theta}) d\omega,
    \label{equ:qtyx}
\end{equation}

Although Equation~\ref{equ:qtyx} provides an analytical solution for the predictive distribution, calculating the integral can be challenging in practical applications. Hence, we employ Monte Carlo sampling \citep{gal2016uncertainty} and series addition method to obtain the final results:
\begin{equation}
    q_{\hat{\theta}}(y \vert x) \approx \frac{1}{N} \sum_{n=1}^{N} p(y \vert x, \hat{\omega}_{n}),
    \label{equ:qmtcl}
\end{equation}
where $N$ denotes the number of samples, and $\hat{\omega}_{n}$ represents the nth sampled value of the weights obtained from $q(\omega \vert \hat{\theta})$. Equation~\ref{equ:qmtcl} is equivalent to Equation~\ref{equ:qtyx} as the number of samples $N$ tends to infinity. This equation indicates us that we could obtain a Bayesian estimation through Monte Carlo sampling of the weights for a predefined neural network.\\

In the study by \citet{PhysRevD.102.103509}, the authors use the total variation principle to derive the analytical results for the variance of the predicted distribution on a fixed input x. They further simplify these results to obtain:
\begin{equation}
    \hat{Var}(y \vert x) \approx \frac{1}{T} \sum_{t=1}^{T} \sigma_t^2 + \frac{1}{T} \sum_{t=1}^{T} (\mu_t^2 - \bar{\mu}^2),
    \label{equ:var}
\end{equation}
In the equation above, $T$ denotes the total number of forward passes of the network. The terms $\sigma_t$ and $\mu_t$ refer to the standard deviation and mean of the distribution obtained during the t-th forward pass, respectively, while $\bar{\mu}$ represents the mean of all $\mu_t$ values. The first term in Equation~\ref{equ:var} corresponds to the aleatoric uncertainty discussed earlier, whereas the second term corresponds to the epistemic uncertainty. In real applications, we build a BPNN with the flipout method discussed in \citet{2018arXiv180304386W} to generate pseudo-independent weight perturbation on minibatches, which could simulate the Bayesian interference process. Then the distribution between predicted results and prior distribution could be evaluated to provide reference to photometry uncertainty. We will discuss details of the method in Section \ref{subsubsection:BPNN}.\\

\subsection{The Structure of Neural Networks in the PNet}\label{subsec:strucNet}
In this subsection, we will provide a comprehensive overview of the neural network structure employed in our framework. The neural network comprises three interconnected components that collaboratively predict the final outcomes. The first component, known as the Photometry-Detection Net, is designed to detect point-like celestial objects, accurately determine their subpixel positions, and calculate their flux. The second component is responsible for estimation of the photometry results and their uncertainties. Lastly we carry out the photometry calibration, utilizing reference stars to ensure precise calibration of the photometric measurements derived from the flux values. This framework allows us to obtain reliable magnitudes and positions for point-like celestial objects. It is important to emphasize that the primary focus of this paper is not the detection of various celestial objects such as galaxies or quasars. Neural networks specifically tailored for the detection of celestial objects from multicolor images are more suitable for these targets \citep{jia2023deep}. Once these targets have been detected by aformentioned methods, we proceed to mask them and carry out point-like celestial object detection, astrometry, and photometry using single-band images with our framework. As a result, the final output of our framework includes the positions of point-like celestial objects, their corresponding magnitudes, and the associated photometry uncertainties, which could be directly used to analyze celestial objects with light variation and moving celestial objects.\\

\subsubsection{The Photometry-Detection Neural Network} \label{subsubsec:PDnet}
Since point-like celestial objects possess a relatively simple structure and smaller size compared to other natural images, there is no need to employ highly complex backbone neural networks specifically designed for natural image target detection. Furthermore, our goal is to determine the center of point-like celestial objects instead of utilizing the bounding box approach commonly used in deep neural network-based target detection algorithms. Therefore, we must modify the position regression strategy within the neural networks. With these considerations in mind, we propose a novel structure known as the Photometry-Detection Net, which directly provides the position and flux of point-like celestial objects. Training data for the Photometry-Detection Net consists of original observation images in a single band, and the corresponding labels in the training data indicate the position and flux of these point-like celestial objects.\\

\begin{figure}
    \plotone{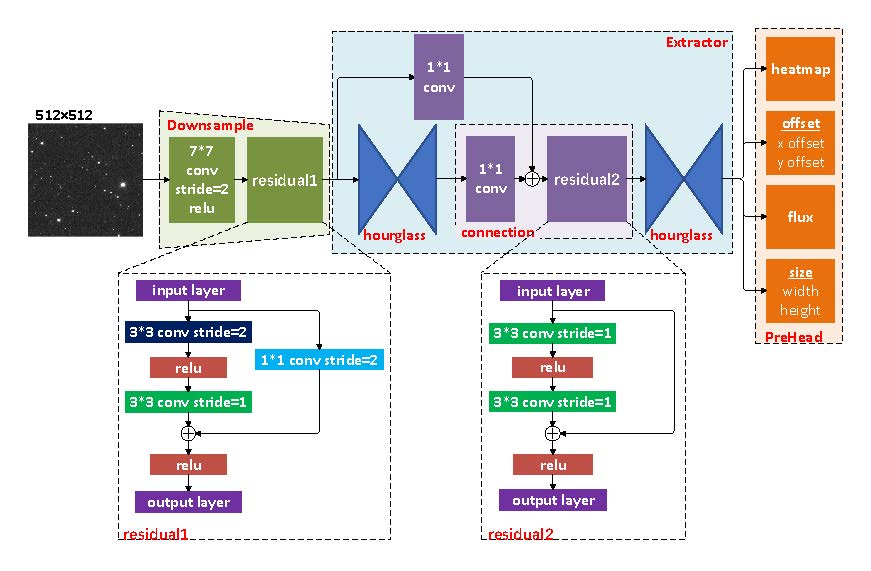} 
    \caption{The structure of CenterNet. The CenterNet includes three modules: the Downsample module, the Extractor module, and  the PreHead module. Residual1 and Residual2 represent two types of residual structures. The network takes an image with a resolution of $512 \times 512$ pixels as input and outputs the center position, flux value, size of the target, and a heatmap that corresponds to the input image. The center position and the flux value will be used for further processing. In the Extractor module, we propose to use the Hourglass neural network for feature extraction.}
    \label{fig:centernet}
\end{figure}

First and foremost, the Photometry-Detection Net integrates the CenterNet structure for the detection of celestial objects. CenterNet is a straightforward, efficient and accurate neural network designed specifically to detect small targets by regressing their key points \citep{2019arXiv190407850Z}. The CenterNet has gained widespread recognition and has been applied in various domains \citep{AHMED2021107489,2021PatRe.11207787G}. The structure of the CenterNet is depicted in Figure~\ref{fig:centernet}. In practical scenarios, the CenterNet begins by performing regression to identify the center point of a target, followed by feature extraction in the vicinity of the center point. Subsequently, the CenterNet provides the target's position and confidence score. Building upon the detection results, we introduce the photometry neural network branch connected to the CenterNet, enabling us to obtain the flux of the point-like celestial objects. To optimize computational resources, we set the input size of the CenterNet to $512 \times 512$ pixels, and the output of the CenterNet consists of the detected targets, their corresponding positions, and their flux values.\\

As depicted in Figure~\ref{fig:centernet}, the CenterNet comprises three distinct modules: the Downsample module, the Extractor module, and the PreHead module. In this study, we employ a 2D convolutional neural network as the Downsample module, using a convolutional layer with a kernel size of $7 \times 7$, a padding of 3, and a stride of 2 to downsample the image to a size of $256 \times 256$ pixels. Subsequently, the resulting image is further downsampled to $128 \times 128$ pixels using the residual1 component, as illustrated in Figure~\ref{fig:centernet}. These down-sampled images are then passed through the Extractor module to extract their features \citep{2014arXiv1412.6806S}. For Extractor module, we adopt the Hourglass neural network in this paper, as it is well-suited for capturing features from small objects. When the downsampled images are inputted into the hourglass network, a sequence of residual modules with convolutional layers having a stride of 2 is employed to iteratively downsample the image four times, reducing its size by a factor of two each time. Following downsampling, four upsampling operations are performed using the nearest-neighbor interpolation algorithm. The spatial information from each downsampling scale is preserved through skip connections and fused with the upsampled feature maps during the upsampling process \citep{10.1007/978-3-319-46484-8_29}. The hourglass network ultimately produces a feature map of the same size as the input. By stacking multiple hourglasses, the detection capability of the neural network can be enhanced, as subsequent hourglasses refine the detection results based on the output of the previous ones, which proves more effective than employing a single detection network. For instance, in a system containing multiple stars, a star that has been missed in the initial detection round may become easier to detect in subsequent hourglasses.\\

\begin{figure}[ht!]
    \plotone{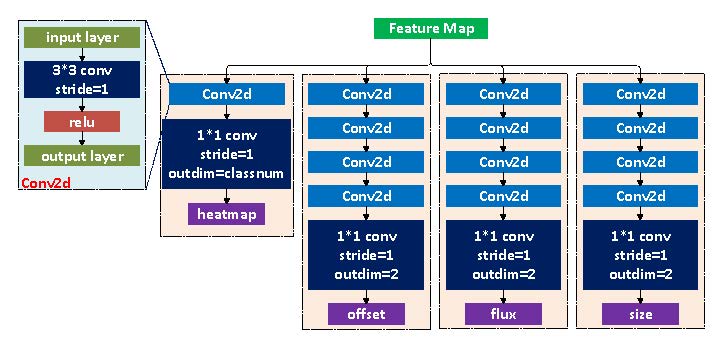}
    \caption{The figure illustrates the structure diagram of the PreHeads. The blue blocks share the same structure, and their detailed configuration is presented in the figure. The yellow 1x1 convolution within the PreHead module is used to adjust the number of channels to align with the desired output parameter count. The input to the PreHead is the feature map extracted by the Hourglass, while the output consists of the parameters indicated in the purple blocks of the figure.}
    \label{fig:prehead}
\end{figure}

Finally, the feature maps extracted from the Extractor module are passed to the PreHead module to perform regression on various attributes of the targets in the image under processing. In this study, the PreHead module consists of four sub-PreHeads: heatmap, offset, flux, and size. All four PreHeads are Convolutional Neural Networks (CNNs) comprising 2D convolutional layers and Rectified Linear Unit (ReLU) layers. The heatmap PreHead generates a heatmap of size $batchsize \times class num \times height \times width$. In this particular study, only star detection is considered, so the class num is set to 1. The heatmap divides the original image into a grid of $128 \times 128$ patches, serving two purposes: providing confidence scores for different classes at various positions of the target and indicating the approximate position of the detected targets within the grid. The offset and size PreHeads predict different properties of the targets but share the same CNN structure, producing output sizes of $batchsize \times 2 \times width \times height$, where 2 represents the two parameters predicted by each PreHead. The Offset PreHead estimates the deviation between the actual center position of the target and the position indicated by the heatmap, which allows us to calculate the actual center position. The Size PreHead provides the height and width of the target. The Flux PreHead has a structure similar to the Offset PreHead, but with a reduced number of output channels (1) and prediction of a single parameter, which represents the flux. The structures of the four PreHeads are depicted in Figure \ref{fig:prehead}. Additionally, it is worth noting that we have incorporated several intermediate supervision steps within the PreHead module. These supervisions are added directly after each hourglass neural network to assess its performance during the training phase. After training, these neural networks will not be utilized, and we will solely employ the main structure of the PreHead module.\\

\subsubsection{The Bayesian Photometry Neural Network}\label{subsubsection:BPNN}
\begin{figure}[ht!]
    \plotone{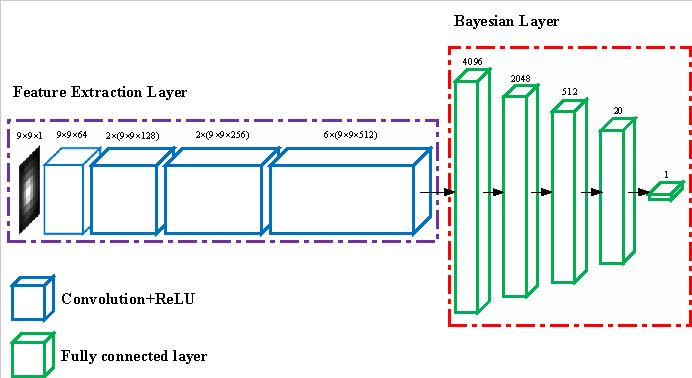}
    \caption{The figure illustrates the architecture of BPNN. The Feature Extraction Layer is depicted in purple, while the Bayesian Layer is represented in red. The blue blocks indicate the convolutional layer, while the green blocks correspond to the Bayesian fully connected layer. The network takes a star image of dimensions $9 \times 9$ pixels as input and produces a predicted distribution for photometry as the output.}
    \label{fig:BPNN}
\end{figure}

The Bayesian Photometry Neural Network (BPNN) is employed to determine the magnitude and uncertainty of photometry results, serving as a reference for subsequent light curve classification. Based on the detection results from the Photometry-Detection Net, all detection results are cropped into stamp images with a size of $9\times 9$ pixels. This size is suitable for star images of moderate brightness, but the stamp images can be adjusted to smaller or larger sizes depending on the actual observation conditions. The stamp images are then used as input for the BPNN, which estimates the magnitude multiple times using the 'flip-out' method to approximate the posterior distribution of the neural networks in magnitude estimation. The BPNN comprises two main components: the Feature Extraction Layer and the Bayesian Layer, as illustrated in Figure~\ref{fig:BPNN}.

The Feature Extraction Layer is built upon the resnet50 architecture \citep{He_2016_CVPR}. Resnet50 tackles the challenges of increased computational time and diminished accuracy in deep network structures by incorporating "shortcut connections". Since the input image is relatively small (9×9 pixels), we have adjusted the size of the convolution kernel to $3\times3$ to better accommodate these small-scale images. Additionally, we have removed the pooling layers in the network to reduce information loss and improve prediction accuracy. Lastly, we have eliminated the fully connected layer and repurposed it as the Feature Extraction Layer.\\

In Section~\ref{subsec:prinbnn}, we have demonstrated that the Variational Bayesian (VB) method can approximate the posterior distribution by minimizing the Evidence Lower Bound (ELBO) as depicted in Equation~\ref{equ:FDtheta}. Typically, this procedure is accomplished using the gradient descent method. The gradient of $F(D,\theta)$ in Equation~\ref{equ:FDtheta} can be computed by considering the density function $q(\omega \vert \theta)$, which is also parameterized by $\theta$:
\begin{equation}
    \nabla_{\theta} E_{q(\omega \vert \theta)} [f_{\theta}(\omega)] = \int \nabla_{\theta} q(\omega \vert \theta) f_{\theta}(\omega) d\omega + E_{q(\omega \vert \theta)} [\nabla_{\theta} f_{\theta}(\omega)].
    \label{equ:diffthe}
\end{equation}
The specific form of $f_\theta(\omega)$ is:
\begin{equation}
    f_\theta(\omega) = \ln{q(\omega \vert \theta)} - \ln{p(D, \omega)}.
    \label{equ:ftomega}
\end{equation}

As depicted in the first term of Equation~\ref{equ:diffthe}, computing the gradient necessitates obtaining analytical solutions for expectations involving the approximate posterior distribution. However, this task often proves challenging in real-world applications. Consequently, when attempting to directly compute Equation~\ref{equ:FDtheta} using a neural network via forward propagation, calculating the gradient becomes generally infeasible, and the computation process lacks differentiability, impeding backpropagation. To address this issue, "The reparameterization trick" has been introduced by \citet{2013arXiv1312.6114K}. By employing this trick, a straightforward and differentiable unbiased estimator for the ELBO can be generated, enabling the use of gradient descent algorithms for ELBO optimization.\\

However, the reparameterization trick is not without its limitations. One issue arises from the fact that all sample weights within the same batch are identical, resulting in correlated gradients across different examples in the batch \citep{PhysRevD.102.103509}. To address this problem, a technique known as "flipout" was introduced by \citet{2018arXiv180304386W}. Flipout facilitates the generation of efficient pseudo-independent weight perturbations on mini-batches. In a comparative study conducted by \citet{PhysRevD.102.103509}, several methods, including Dropout, Dropconnect, Reparameterization Trick (RT), and flipout, have been evaluated, with flipout demonstrating superior performance. As a result, for the implementation of the Bayesian Neural Network in the BPNN, we have opted to utilize flipout. Specifically, we employ the method proposed in \citet{ranganath_krishnan_2022_5908307} to construct a 5-layer flipout linear layer, thereby enabling the realization of Bayesian layers. In the flipout linear layers, multiple sets of weights are sampled during each forward pass. During the training step, the weights are randomly flipped or flipped out based on a Bernoulli distribution. In the deployment step, these weights are treated as random variables and can be used for uncertainty estimation. By sampling multiple sets of weights, flipout introduces stochasticity into the network, resulting in different predictions for the same input. This characteristic allows for the quantification of uncertainty and enhances the model's robustness and generalization capabilities.\\

\subsubsection{The Photometry Calibration Part}\label{subsubsection:PhotometryCalibration}
With above neural networks, we are able to estimate the flux of celestial objects based solely on the grayscale counts within the images of these objects. However, in order to identify flares and variable stars, it becomes crucial to calibrate the flux of all detected targets by referencing the flux from reference stars. For this calibration process, we employ the method proposed by \citet{Stoughton_2002} to convert flux to magnitudes, which is represented by the Equation~\ref{equ:magcal}:
\begin{equation}
    mag = mag_{0} + mag_{zero} + k(t) \times x + f(i)
    \label{equ:magcal}
\end{equation}

The equation provided calculates the calibrated magnitude $mag$ based on various variables. The instrumental magnitude $mag_{0}$ is derived from the photometry branch, and $mag_{zero}$ represents the zero-point magnitude set by the user. The primary extinction coefficient is denoted as $k$, and $x$ represents the air mass. Additionally, $f(i)$ characterizes the flat field of the CCD's $i$th column. Consequently, when working with SDSS data, it is necessary to obtain the run, camcol, and field parameters for magnitude calibration. On the other hand, for data obtained by different telescopes, we need to firstly use the PNet to obtain flux and detection results. Then, we use 50 to 100 isolated reference stars on the same image with known magnitudes to fit the calibration function by either least squares or minimizing the chi-square distance between the flux and the magnitudes. In the above steps, we will filter out abnormal points where the error between the observed value and the expected value exceeds $1\sigma$.\\
\section{Training and Performance Evaluation} \label{sec:tpe}
In this chapter, we will utilize two types of data to train and evaluate the performance of our framework: the g-band data from SDSS DR17 and simulated data generated by Skymaker. It is important to highlight that while the datasets differ, all other computational procedures in the framework remain consistent. Besides, we select the SExtractor\citep{bertin1996sextractor}, a commonly used astrometry and photometry tool for data processing, to process the observation data at the same time for comparison. As a benchmark, we will employ the magnitudes and coordinates provided by the official catalogue from the SDSS for real observation data and catalog provided by the Skymaker for simulated data. Regarding the Photometry-Detection Net, we will showcase the detection results as well as preliminary photometry results. Subsequently, we will focus on the Bayesian Photometry Neural Network and present the estimation of the photometry uncertainty.

\subsection{Training and Performance Evaluation of the Photometry-Detection Net}\label{PDNRes}

During the training phase, certain targets that are not utilized in the training data, such as dense star fields and stars close to galaxies, are masked. As previously mentioned, the detection of these targets from multiple color images requires specialized processing methods, which are beyond the scope of this paper. For optimization, we employ the Adam Optimizer \citep{Kingma2014AdamAM} and evaluate the classification results using the Focal Loss \citep{Lin_2017_ICCV}.
The Focal Loss dynamically adjusts the impact of each training example on the overall loss, ensuring a balanced approach to the detection results of celestial objects with varying magnitudes. The Focal Loss is formulated by introducing a modulating factor into the Cross Entropy (CE) loss, as originally proposed by \citet{Lin_2017_ICCV}. In this study, we employ an alpha-balanced version of the focal loss, as defined in Equation~\ref{equ:focalloss}:
\begin{equation}
    Focal Loss(p_t) = -\alpha_t (1-p_t)^\gamma log(p_t),
    \label{equ:focalloss}
\end{equation}
where $\alpha_t$ is the weighting factor used to adjust the weights of positive and negative samples, $p_t$ is the output of the network, and $\gamma > 0$ is an adjustable focusing parameter (2 in this paper). The Mean Absolute Error (MAE) loss is utilized to assess position error, while the Mean Square Error (MSE) loss is employed to evaluate flux prediction outcomes. The computation of MSE and MAE is demonstrated in Equation~\ref{equ:msae}, where $n$ represents the total number of targets, $\hat{y}_i$ denotes the predicted value of the i-th target, and $y_i$ denotes the corresponding true value. The aggregate of these three losses constitutes the loss function employed to train the Photometry-Detection Net, as depicted in Equation~\ref{equ:msae}:
\begin{equation}
    \begin{aligned}
        MAE &= \frac{1}{n} \sum_{i=1}^n \lvert \hat{y}_i - y_i \rvert \\
        MSE &= \frac{1}{n} \sum_{i=1}^n (\hat{y}_i - y_i)^2.
    \end{aligned}
    \label{equ:msae}
\end{equation}
It takes approximately 324 seconds to complete a single epoch when training the Photometry-Detection Net on a computer equipped with a 3090Ti GPU. The batch size consists of 40 images, each with dimensions of $512\times 512$ pixels. After approximately 40 to 50 epochs of training with randomly initialized weights, the Photometry-Detection Net starts to converge. The convergence process is accelerated when using pre-trained weights. Since the instrumental magnitude measurements exhibit a relatively consistent pattern, when employing pre-trained weights and training on new data batches, it is common practice to freeze the feature extraction network (the Hourglass Network) and the photometry branch network. After training for a specific number of epochs, these two networks are then unfrozen, and the overall network undergoes fine-tuning.\\

\begin{figure}[ht!]
    \centering
    \subfigure[The relation between the confidence and the precision and the recall for the Photometry-Detection Net in the SDSS data.]{\includegraphics[width=0.4\textwidth]{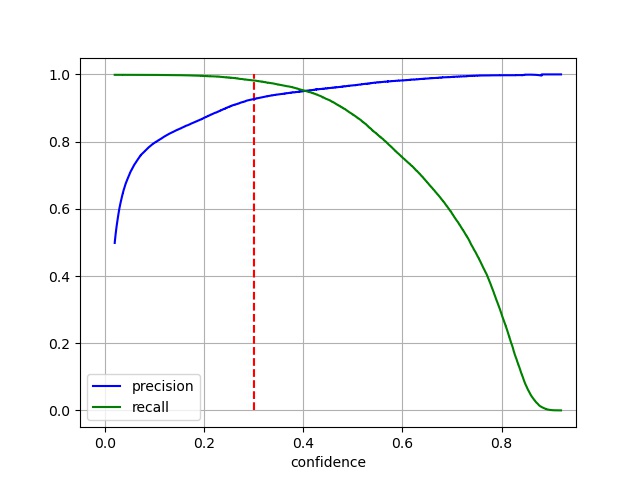}\label{fig:SDSScon}}
    \hspace{1.5cm}
    \subfigure[The relation between the confidence and the precision and the recall for the Photometry-Detection Net in the simulated data.]{\includegraphics[width=0.4\textwidth]{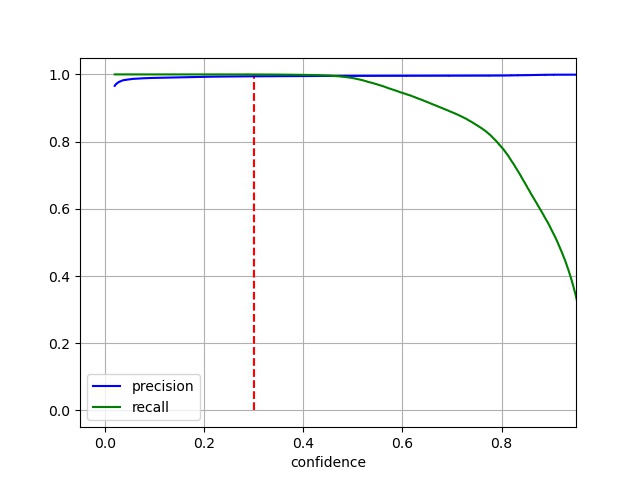}\label{fig:simcon}}
    \vspace{0.5cm} 
    \subfigure[The precision-recall curve for the Photometry-Detection Net in the SDSS data.]{\includegraphics[width=0.4\textwidth]{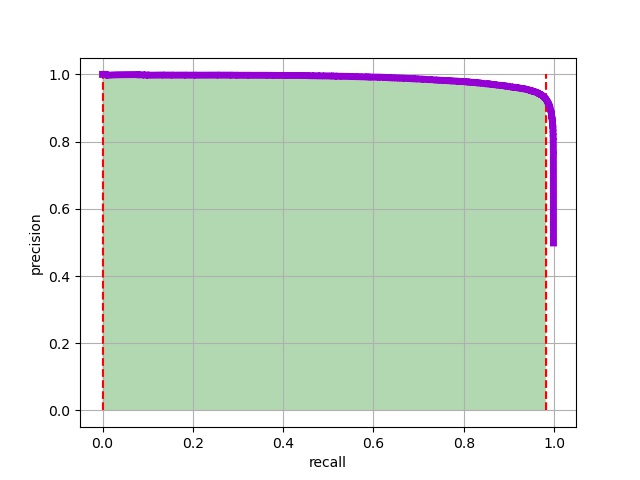}\label{fig:SDSS}}
    \hspace{1.5cm}
    \subfigure[The precision-recall curve for the Photometry-Detection Net in the Simulated data.]{\includegraphics[width=0.4\textwidth]{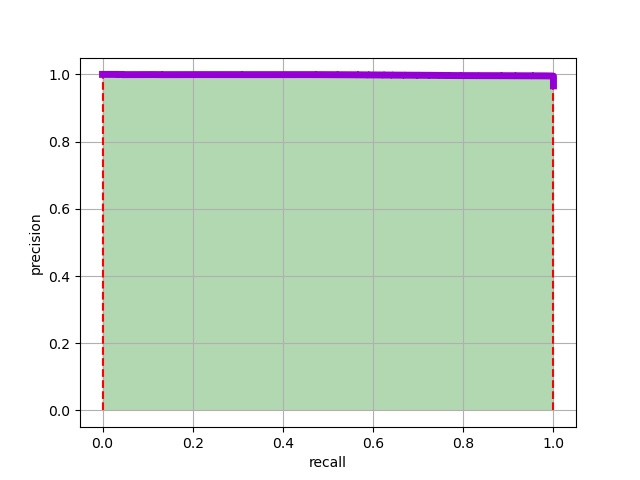}\label{fig:sim}} 
    \caption {In figures a and b, the horizontal axis represents the confidence level of the predicted targets by the Photometry-Detection Net, while the vertical axis represents the corresponding precision and recall at that confidence level. The red dashed line represents the confidence threshold of 0.3 that has been used in this study. In figures c and d, the horizontal axis represents recall and the vertical axis represents precision. The purple curve illustrates the precision-recall (p-r) curve when the confidence threshold is set to 0. The area between the two red dashed lines depicts the p-r curve when the confidence threshold is set to 0.3, and the region between this curve and the horizontal axis is visually represented as the green area in the figure. This provides insights into the Photometry-Detection Net's performance in detecting star targets. As explained in this section, the area between the p-r curve and the horizontal axis corresponds to the average precision (AP). A higher AP value indicates a better target detection performance by the network. As shown in this figure, our method could obtain AP more than $98\%$ in simulated and in real observation data.} 
    \label{fig:sdsssimRes}
\end{figure}

After training, we have tested the performance of Photometry-Detection Net in detecting star targets and measuring magnitues using a batch of $512 \times 512$ pixel SDSS astronomical images and simulated images generated by Skymaker. When the network outputs classification results, it provides a confidence level, and when this level is higher than a certain threshold, we consider that there exist star targets. Generally, the lower the threshold, the higher the recall, but the lower the precision, and vice versa. We have shown the performance of Photometry-Detection Net using different confidence thresholds, as shown in Figure~\ref{fig:sdsssimRes}. When the confidence threshold is set to 0.3, the specific results obtained from testing on SDSS data and simulated data are shown in Table~\ref{table:prm}. The table also presents the detection results from SExtractor. It's evident that our method achieves a higher level of precision and recall rate compared to SExtractor. 

\begin{table}[h]
    \setlength{\abovecaptionskip}{0.05cm}
    \centering
    \caption{Detection Results}
    \begin{tabular*}{0.6\hsize}{@{\extracolsep{\fill}}c c c c}
      \toprule\toprule 
       & SExtractor\_SDSS & PNet\_SDSS & PNet\_Simulation\\
      \midrule 
      Precision & $84.24\%$ & $92.64\%$ & $99.43\%$ \\
      Recall & $96.12\%$ & $98.20\%$ & $99.98\%$ \\
      % mAP & $98.56\%$ & $99.83\%$ \\
      \bottomrule 
    \end{tabular*}
    \label{table:prm}
\end{table}

%Table~\ref{table:prm} provides the mAP values for the Photometry-Detection Net, indicating $99.83\%$ for simulated data and $98.56\%$ for real observation data, which meets the requirements of our study. However, it should be noted that these values may vary for different astronomical image datasets and should be fine-tuned through testing. By carefully selecting an appropriate confidence threshold to balance precision and recall, the Photometry-Detection Net can achieve satisfactory detection performance tailored to specific application needs.\\

\begin{figure}[ht!]
    \centering
    \subfigure[The Photometry error of the SExtractor for the SDSS Data]{\includegraphics[width=0.4\textwidth]{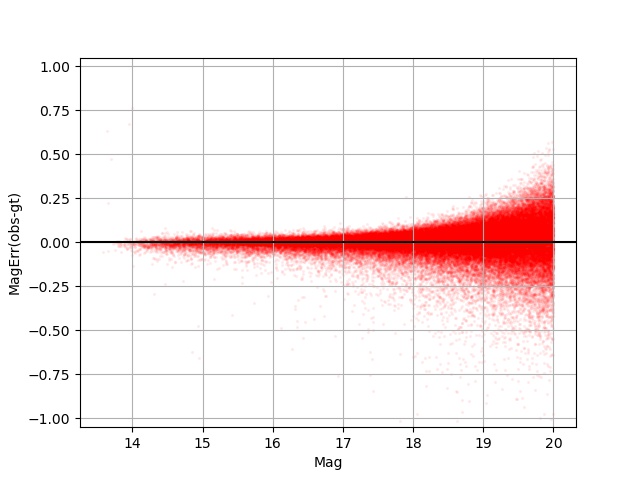}\label{fig:SDSSpho}}
    \hspace{1.cm}
    \subfigure[Histogram of the Photometry error of the SExtractor for the SDSS Data]{\includegraphics[width=0.5\textwidth]{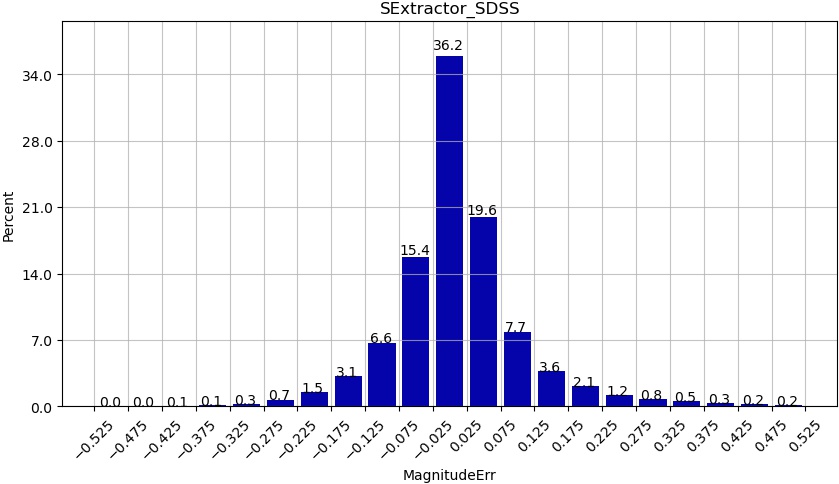}\label{fig:SDSSphosta}}
    \subfigure[The Photometry error of the PNet for the SDSS Data]{\includegraphics[width=0.4\textwidth]{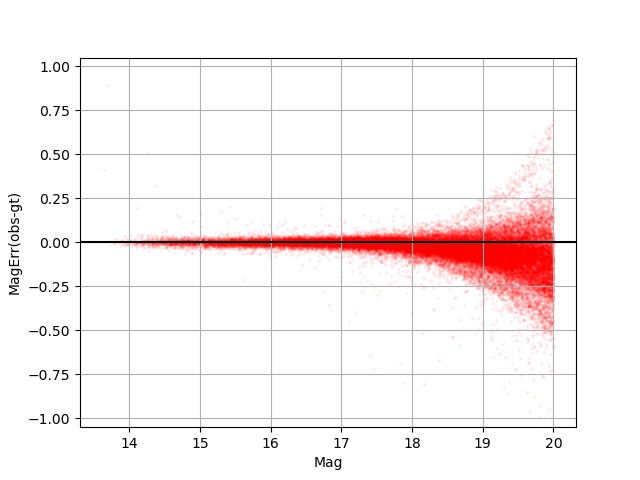}\label{fig:simpho}}
    \hspace{1.cm}
    \subfigure[Histogram of the Photometry error of the PNet for the SDSS Data]{\includegraphics[width=0.5\textwidth]{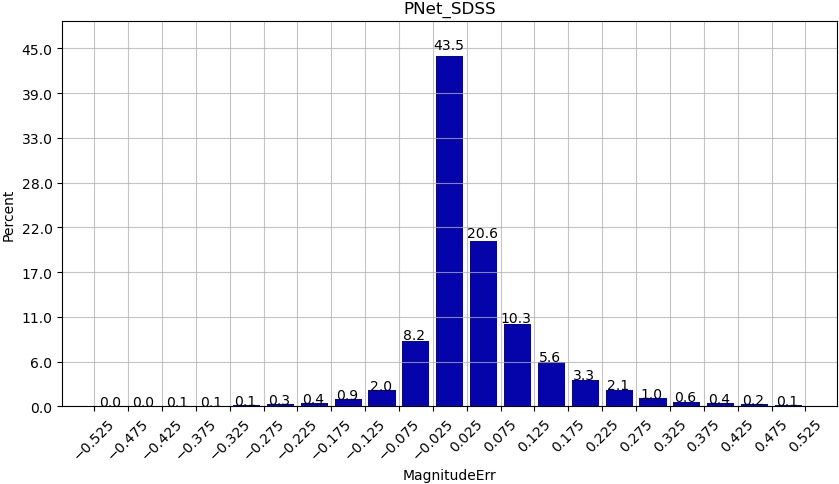}\label{fig:SDSSphosta}}
    \subfigure[The Photometry error of the PNet for the Simulated Data]{\includegraphics[width=0.4\textwidth]{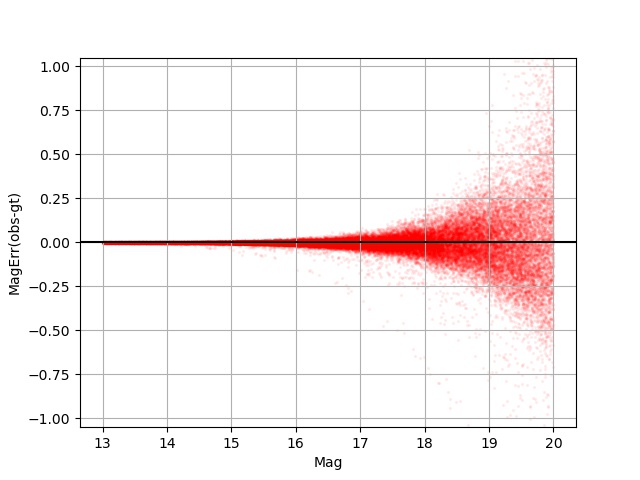}\label{fig:simpho}}
    \hspace{1.cm}
    \subfigure[Histogram of the Photometry error of the PNet for the Simulated Data]{\includegraphics[width=0.5\textwidth]{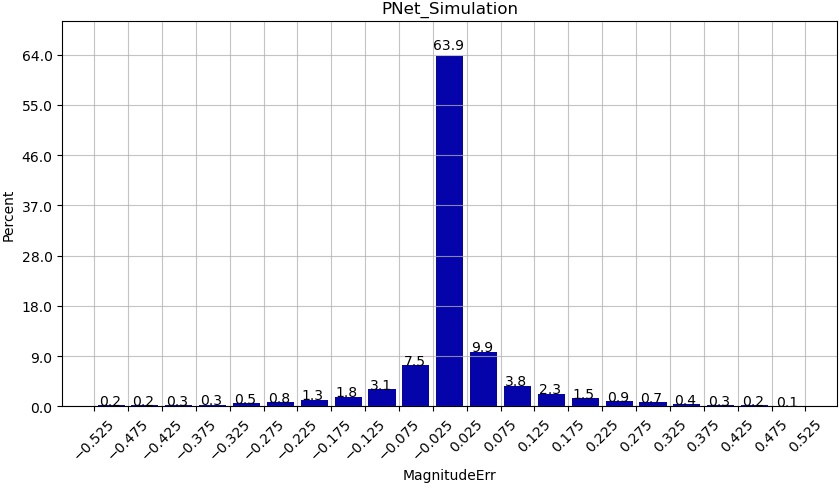}\label{fig:simphosta}} 
    \caption{Figure a, c and e display the photometric error of the SExtractor on SDSS data, the Photometry-Detection on SDSS data, and the Photometry-Detection on Simulated data respectively. In the top figure, the x-axis represents the magnitude, while the y-axis represents the error between the predicted and true magnitudes. The proximity of the scatter points to the y=0 line indicates the accuracy of the network's photometry results. The figure illustrates that our algorithm exhibits smaller errors in regions with lower magnitudes compared to those with higher magnitudes. Figures b, d, and f display histograms of magnitude errors. The horizontal axis signifies the magnitude error range, while the vertical axis indicates the percentage of stars falling into each error range. A larger concentration of stars within smaller magnitude error ranges on the horizontal axis indicates a more robust capability of the algorithm to measure target magnitudes. As depicted in these figures, the PNet exhibits superior photometry results compared to those of SExtractor.}
    \label{fig:sdsssimpho}
\end{figure}

\begin{figure}[ht!]
    % \plotone{pic/PDNetCoor.jpg}
    \centering
    \includegraphics[width=0.95\textwidth]{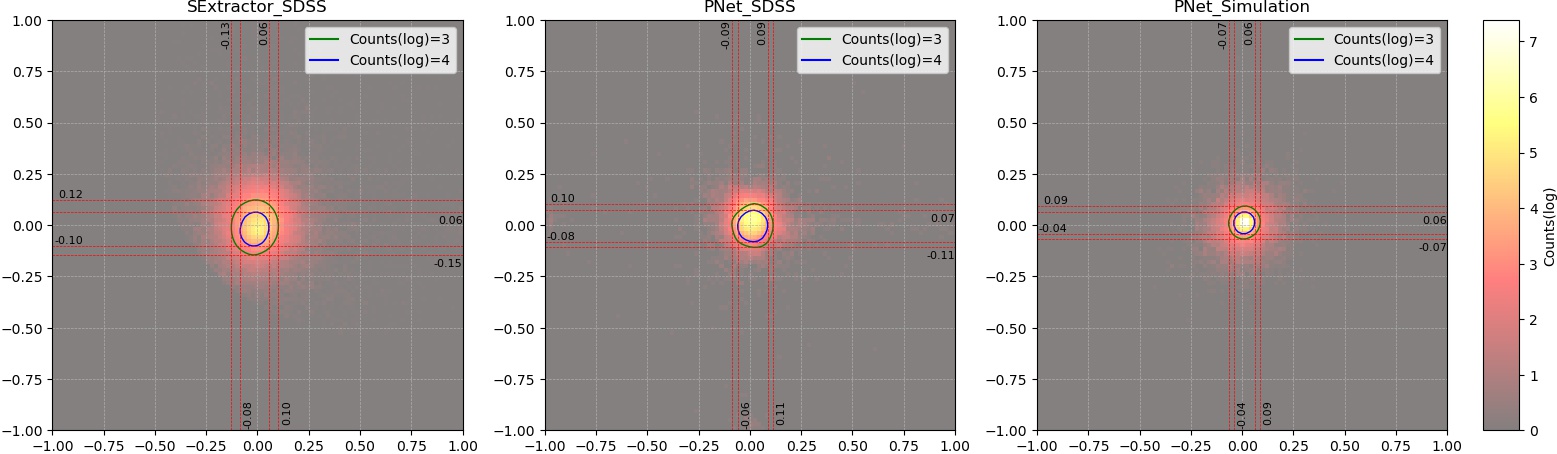}
    \caption{The left panel displays the astrometry error of SExtractor on the SDSS data, the middle panel illustrates the astrometry error of the Photometry-Detection Net on the SDSS data, and the right panel showcases the astrometry error of the Photometry-Detection Net on the simulated data. Assuming the true positions of all targets are centered at (0,0), the predicted positions are scattered across the figure. Each small area within the figure, defined as a $0.02 \times 0.02$ pixel region, tallies the number of stars with astrometry errors falling within that area. The resulting count is logarithmically transformed and visualized as a heatmap. Both the horizontal and vertical axes are measured in pixels, while the color bar denotes the logarithm of the star count. The blue contour corresponds to an astrometry error of 0.15 pixels, the green contour corresponds to an astrometry error of 0.21 pixels, and the red dashed line represents the limit of astrometry error in both the x and y directions. When the centers of the green and blue contour lines are closer to the (0,0) point, and the contours are smaller and more symmetrical, it indicates improved astrometry accuracy in the algorithm.}
    \label{fig:PDNetCoor}
\end{figure}

During the star target detection process, the Photometry-Detection Net also conducts photometry and astrometry. For the SDSS data, the photometry branch utilizes the calibration method described in Section~\ref{subsubsection:PhotometryCalibration} to calibrate the magnitudes. However, since the simulated data does not account for airmass effects or other effects, calibration is not necessary for these data. When using SExtractor to process SDSS data, a fixed offset between the photometry and astrometry results and their ground truth values may exist. This offset can be determined by statistically analyzing the measurement results. In our study, we have adjusted the results obtained from SExtractor by subtracting this offset. Aside from this adjustment, no further processing has been performed on the measurement results derived from SExtractor for the SDSS data. The photometry and astrometry results are depicted in Figure~\ref{fig:sdsssimpho} and Figure~\ref{fig:PDNetCoor}.\\ 

Figure~\ref{fig:sdsssimpho} reveals that the magnitude error is mostly within 0.5 magnitudes for the majority of stars, with approximately $50\%$ of the stars exhibiting magnitude measurement errors within $2.5\%$. These measurement errors tend to increase gradually as the magnitude of the stars increases. This behavior aligns with theoretical analysis, as the impact of noise with the same level has a smaller effect on stars with lower magnitudes. For every 5-magnitude decrease, the flux of the star increases by a factor of 100. Additionally, compared to the errors observed in the SDSS data, the measurement errors of star magnitudes in the simulated data are generally smaller and more concentrated. This difference primarily stems from the presence of noise in real data, which not only interferes with the algorithm's flux measurements but also affects the process of calibrating star magnitudes. Additionally, as illustrated in Figure 7, it's worth noting that the photometric error of our framework when applied to SDSS data exhibits some asymmetry, primarily due to the calibration process. However, when looking at the overall performance, our framework shows more consistent and reliable photometric results, outperforming those obtained by SExtractor on the SDSS data.\\

Meanwhile, we compared the astrometry results obtained by the Photometry-Detection Net and SExtractor in Figure~\ref{fig:PDNetCoor}. The astrometry error manifests as a roughly symmetrical circle. The contour lines marked with values 3 represent a diameter of 0.15 pixels, while those with values 4 represent a diameter of 0.21 pixels. In contrast to SExtractor, our framework demonstrates superior astrometry accuracy when applied to the SDSS data, with a higher number of stars exhibiting astrometry error smaller than 0.1. Nevertheless, it's worth noting that the astrometry error for the SDSS data remains somewhat larger than that observed in the simulated data, primarily due to the introduction of other noise during real observations.\\

\subsection{Training and Performance Evaluation of the Bayesian Photometry Neural Network}\label{BPNNRes}
As mentioned earlier, the detected results in the images will be segmented into smaller patches to estimate photometry results and uncertainties. We assume the Gaussian distribution as the prior distribution for the photometry results, and the loss function can be derived directly from Equation~\ref{equ:FDtheta}. For optimization during the training stage, we utilize the Adam optimizer \citep{Kingma2014AdamAM}. However, during the training stage, there is a possibility of encountering the issue of exploding gradients due to the random weight sampling, which can lead to unstable training stage. To address this problem, we implement gradient clipping by setting a threshold. If the gradient surpasses this threshold, it is truncated to a specific value. On average, training one epoch typically takes approximately 12 seconds for a batch of 2000 small images.\\

\begin{figure}[ht!]
    \centering
    \subfigure[SDSS photometry results and uncertainties obtained by the Baysian Neural Network]{\includegraphics[width=0.8\textwidth]{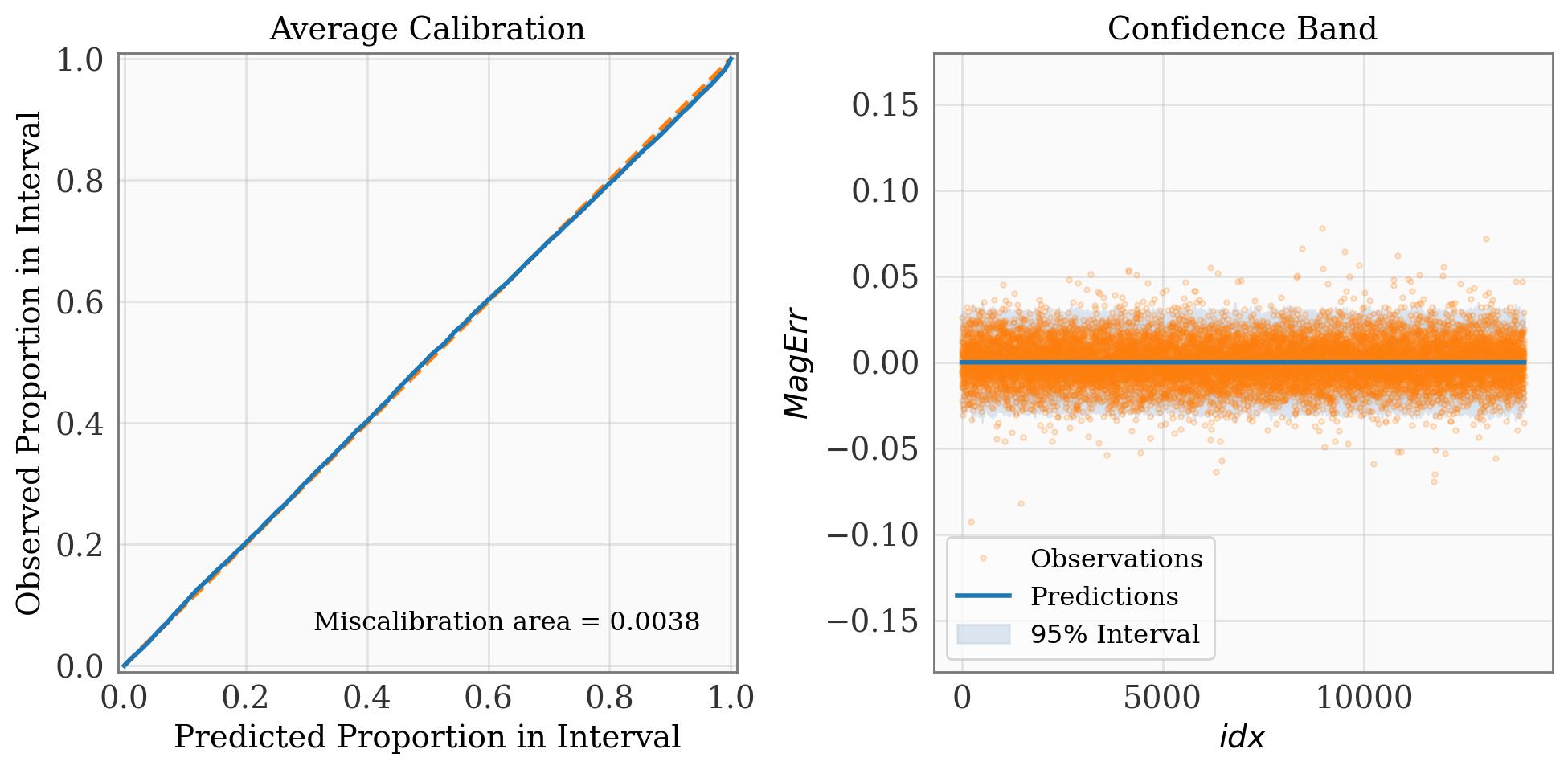}\label{fig:SDSSunc}}
    \subfigure[Simulated photometry results and uncertainties obtained by the Baysian Neural Network]{\includegraphics[width=0.8\textwidth]{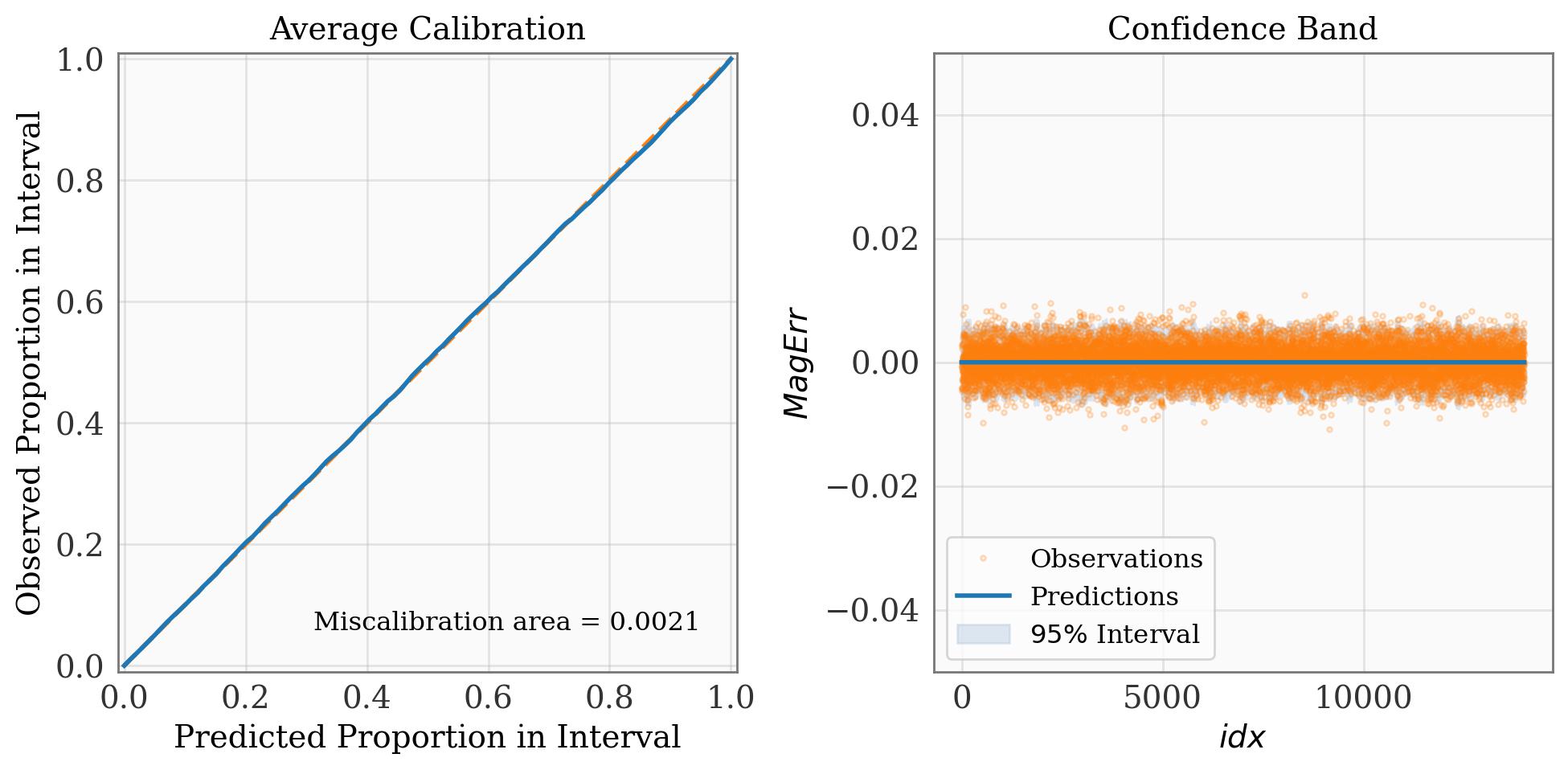}\label{fig:simunc}}
    \caption{Figure a displays the outcomes of the Bayesian Photometry Neural Network in measuring the photometry of targets within the SDSS data, whereas figure b showcases the results for simulated data. In the "Average Calibration" plot \citep{tran2020methods}, the horizontal axis represents the expected proportion, while the vertical axis represents the actual observed proportion. The "Miscalibration area" pertains to the space between the curve and the dashed diagonal line with a slope of one. A smaller value indicates more accurate uncertainty calibration results. In the "Confidence Band" plot, the horizontal axis corresponds to the target index, and the vertical axis represents the error between the mean of the target distribution calculated by the Bayesian Photometry Neural Network and the true value. To align all mean values of the target distributions with zero, we shift them accordingly, as indicated by the blue line at y = 0 within the figure. The orange dots illustrate the discrepancy between the true values and the mean of the target distribution, while the blue region denotes the $95\%$ confidence interval projected by the Bayesian Photometry Neural Network. A favorable outcome in uncertainty prediction occurs when the light blue region encompasses the orange dots.}
    \label{fig:sdsssimunc}
\end{figure}

Upon completing the training phase, we proceed to assess the performance of the Bayesian Photometry Neural Network using a dataset of 14,000 target samples. To obtain the distribution of estimated magnitudes for each star target, we employ Monte Carlo sampling with 50 discrete samples. It is worth noting that due to the large number of parameters involved and the challenges inherent in optimizing Bayesian Neural Networks, the estimated uncertainty distribution may not closely align with the true distribution. In order to address this, we propose utilizing the method introduced by \citet{chung2021uncertainty} to calibrate the predicted uncertainty results. This calibration algorithm, based on isotonic regression \citep{pmlr-v80-kuleshov18a}, adjusts the uncertainty represented by the standard deviation by determining an appropriate calibration coefficient. After measuring the magnitudes of all star targets, the Bayesian Photometry Neural Network provides corresponding mean values and standard deviations to characterize the uncertainty distribution. We employ Equation~\ref{equ:nordis} to standardize these distributions:
\begin{equation}
    NormalDistribution = \frac{y_{true} - \mu}{\sigma}.
    \label{equ:nordis}
\end{equation}
Here, $\mu$ represents the mean of the predicted distribution, $y_{true}$ represents the true value of the corresponding target, and $\sigma$ denotes the standard deviation of the predicted distribution. By utilizing the aforementioned method, we are able to obtain the distribution of all targets on a standardized evaluation scale. Subsequently, we compare the standardized distribution of magnitudes for each target with a standard normal distribution having a mean of 0 and a standard deviation of 1. This comparison allows us to assess the accuracy of our Bayesian network's quantification of uncertainty. The results are depicted in Figure~\ref{fig:sdsssimunc}.\\

More specifically, we utilize the standard normal distribution as the probability density function (PDF) to conduct a symmetrical search for boundaries from the center outward. This search aims to identify the boundaries where the ratio of the integral of the PDF within the boundary to the infinite integral of the PDF equals the horizontal axis value in the Average Calibration plot of Figure~\ref{fig:sdsssimunc}. The vertical axis represents the ratio of the number of standardized targets within the observed boundary to the total number of targets. Since the PDF follows a standard normal distribution, the infinite integral value is 1. In an ideal scenario where uncertainty can be accurately quantified, these two ratios should align, resulting in a diagonal line with a slope of 1 in the Average Calibration plot of Figure~\ref{fig:sdsssimunc}. When the curve bends downwards, indicating that the Predicted proportion in the interval exceeds the Observed proportion in the interval, it signifies an overconfident state of uncertainty estimation (where the predicted uncertainty is too small). Conversely, when the curve bends upwards, indicating that the Predicted proportion in the interval falls below the Observed proportion in the interval, it suggests an under-confident state of uncertainty estimation (where the predicted uncertainty is too large).\\

Overall, the Bayesian Photometry Neural Network has the ability to predict the uncertainty distribution of magnitude estimation results. Once we have obtained the magnitude distribution predictions for all targets using this network, we can assess these predictions using the evaluation criterion described in Section~\ref{subsec:criterion}. By defining the boundary of the relative error as positive and negative $3 \sigma_{relative}$, we can examine the results presented in Table~\ref{table:uncRes}. It is worth noting that the calculation method for $\sigma_{relative}$ involves initially calculating the relative error of all targets using the procedure outlined in Equation~\ref{equ:nmad2}, and subsequently determining the standard deviation of these relative errors as $\sigma_{relative}$. This calculation method differs from the $\sigma$ value that represents the uncertainty computed from the distribution of an individual target. Based on the data presented in Table~\ref{table:uncRes}, it becomes apparent that the outlier fraction obtained from the simulated data is lower compared to that from the real data, suggesting that the algorithm demonstrates greater stability when applied to simulated data. Furthermore, the $sigma_{NMAD}$ and MAE values, assessed for both simulated and real data, indicate that the dispersion of predictions from this algorithm is lower, resulting in overall smaller errors when dealing with simulated data. Additionally, the parameter $\bar{E}$ indicates that the Bayesian Photometry Neural Network expresses higher confidence in magnitude measurements for simulated data. In general, the results obtained from simulated data exhibit higher accuracy and confidence compared to real data. This discrepancy could potentially be attributed to the effects induced by noise present in real observational data. On one hand, the presence of noisy pixels can interfere with the Resnet50 model's extraction of image features, particularly when stars have low signal-to-noise ratios, intensifying such interference. On the other hand, noise can also disrupt the statistical photometry results obtained by the Bayesian layer when leveraging image information. This interference manifests as increased uncertainty in the output results, consequently indicating lower reliability and credibility.\\
\begin{table}[h]
    \setlength{\abovecaptionskip}{0.05cm}
    \centering
    \caption{Uncertainty of photometry measurement Results on SDSS and Simulated Data}
    \begin{tabular*}{0.7\hsize}{@{\extracolsep{\fill}}c c c}
      \toprule\toprule 
       & SDSS & Simulation\\
      \midrule 
      $\eta$ & $0.3143\%$ & $0.1879\%$ \\
      $\sigma_{NMAD}$ & $6.047 \times 10^{-4}$ & $1.573 \times 10^{-4}$ \\
      MAE & $9.697 \times 10^{-3}$ & $2.105 \times 10^{-3}$ \\
      $\bar{E}$ & $7.004 \times 10^{-2}$ & $1.682 \times 10^{-2}$ \\
      \bottomrule
    \end{tabular*}
    \label{table:uncRes}
\end{table}

\section{Conclusions and future works}\label{ConFuture}
In this paper, we introduce the PNet, a novel approach for star detection, photometry, and estimation of photometry uncertainties. By leveraging the Bayesian Photometry Network and the Photometry-Detection Net, the PNet offers a comprehensive solution for photometry and astrometry of point-like celestial objects. To evaluate its performance, we conduct tests using both SDSS data and simulated data. The results indicate that our algorithm achieves consistent and reliable results in the simulated data. However, when applied to real data, the presence of noise or undisclosed data processing steps may introduce certain errors. Nonetheless, the overall results are deemed satisfactory.\\

There are several additional points that need to be addressed. First, it is crucial to investigate and adopt data preprocessing methods proposed by other teams for magnitude and position calibration. Given the prevalent use of CMOS cameras, which differ significantly from CCD cameras, we must also explore suitable data pre-processing approaches specific to CMOS cameras. Furthermore, it is important to acknowledge the rapid advancements in neural networks in recent years. Exploring alternative methods such as neural network search or meta-learning could potentially yield improved neural network architectures. Lastly, integrating the results obtained from the Bayesian Photometry Neural Network with the light curve classification algorithm is essential for the development of new techniques in transient discovery. These techniques will prove valuable for future sky survey projects, such as the Large Synoptic Survey Telescope(LSST) \citep{eljko2008}, the Chinese Space Station Telescope(CSST) \citep{Gong_2019} and the SiTian Project \citep{liu_soria_wu_wu_shang_2021}.\\

%% Also note that the akcnowlodgment environment does not support long amounts of text. If you have a lot of people and institutions to acknowledge, do not use this command. Instead, create a new \section{Acknowledgments}.
\section*{acknowledgments}
First, we express our gratitude to the reviewer, whose valuable feedback and guidance over the course of more than two years have significantly contributed to the improvement of our method. Peng Jia would like to thank Professor Zhaohui Shang from National Astronomical Observatories, Professor Jian Ge from Shanghai Astronomical Observatory, Professor Rongyu Sun from Purple Mountain Observatory, Professor Huigen Liu from Nanjing University, Professor Chengyuan Li and Professor Bo Ma from Sun Yat-Sen University who provide very helpful suggestions for this paper. Furthermore, we would like to announce that the code utilized in this article will be made available in the PaperData repository, which is powered by China-VO, ensuring easy access for interested researchers.\\

Furthermore, we express our gratitude for the generous financial support provided by the National Natural Science Foundation of China (NSFC) under grant numbers 12173027 and 12173062, as well as the Major Key Project of PCL. We also acknowledge the science research grants received from the China Manned Space Project with NO. CMS-CSST-2021-A01 and the Square Kilometre Array (SKA) Project with NO. 2020SKA0110102. Additionally, we extend our appreciation to the Civil Aerospace Technology Research Project (D050105) and the French National Research Agency (ANR) for their support in the form of the ANR APPLY grant (ANR-19-CE31-0011) coordinated by B. Neichel.\\

Funding for the Sloan Digital Sky Survey IV has been provided by the Alfred P. Sloan Foundation, the U.S. Department of Energy Office of Science, and the Participating Institutions. SDSS-IV acknowledges support and resources from the Center for High Performance Computing at the University of Utah. The SDSS website is www.sdss4.org. SDSS-IV is managed by the Astrophysical Research Consortium for the Participating Institutions of the SDSS Collaboration including the Brazilian Participation Group, the Carnegie Institution for Science, Carnegie Mellon University, Center for Astrophysics | Harvard \& Smithsonian, the Chilean Participation Group, the French Participation Group, Instituto de Astrof\'isica de Canarias, The Johns Hopkins University, Kavli Institute for the Physics and Mathematics of the Universe (IPMU) / University of Tokyo, the Korean Participation Group, Lawrence Berkeley National Laboratory, Leibniz Institut f\"ur Astrophysik Potsdam (AIP),  Max-Planck-Institut f\"ur Astronomie (MPIA Heidelberg), Max-Planck-Institut f\"ur Astrophysik (MPA Garching), Max-Planck-Institut f\"ur Extraterrestrische Physik (MPE), National Astronomical Observatories of China, New Mexico State University, New York University, University of Notre Dame, Observat\'ario Nacional / MCTI, The Ohio State University, Pennsylvania State University, Shanghai Astronomical Observatory, United Kingdom Participation Group, Universidad Nacional Aut\'onoma de M\'exico, University of Arizona, University of Colorado Boulder, University of Oxford, University of Portsmouth, University of Utah, University of Virginia, University of Washington, University of Wisconsin, Vanderbilt University, and Yale University.

\software{Skymaker \citep{bertin2009skymaker},  
          Bayesian-Torch \citep{ranganath_krishnan_2022_5908307}, 
          Uncertainty Toolbox \citep{chung2021uncertainty},
          PyTorch \citep{paszke2019pytorch},
          Astropy \citep{robitaille2013astropy},
          Matplotlib \citep{hunter2007matplotlib},
          Numpy \citep{harris2020array},
          Scipy \citep{virtanen2020scipy},
          Pandas \citep{mckinney2011pandas},
          tqdm \citep{da2019tqdm},
          photutils \citep{2016asclsoft09011B},
          OpenCV \citep{bradski2008learning},
          Pillow \citep{pillow}
          }

\bibliography{sample631}{}
\bibliographystyle{aasjournal}



\end{document}